\documentclass{iopart}
\usepackage{graphicx}
\usepackage{hyperref}
\usepackage{color}
\usepackage[normalem]{ulem}
\newlength{\deftabcolsep}
\begin{document}
\title[Mechanism of metallization and superconductivity suppression \ldots
by $^{67}$Zn NQR]{Mechanism of metallization and superconductivity suppression
in YBa$_2($Cu$_{0.97}$Zn$_{0.03})_3$O$_{6.92}$ revealed by $^{67}$Zn NQR}
\author{D.~Pelc, M.~{Po\v zek}, V. Despoja, and D.~K.~Sunko$^1$}
\address{Department of Physics, Faculty of Science, University of Zagreb,\\
Bijeni\v cka cesta 32, HR-10000 Zagreb, Croatia.}
\ead{$^1$dks@phy.hr}
%\date{}
\newcommand{\ReS}{\mathrm{Re}\,\Sigma}
\newcommand{\ImS}{\mathrm{Im}\,\Sigma}
\newcommand{\eV}{\;\mathrm{eV}}

\begin{abstract}

We measure the nuclear quadrupole resonance (NQR) signal on the Zn site in
nearly optimally doped YBa$_2$Cu$_3$O$_{6.92}$, when Cu is substituted by 3\%
of isotopically pure $^{67}$Zn. We observe that Zn creates large insulating
islands, confirming two earlier conjectures: that doping provokes an orbital
transition in the CuO$_2$ plane, which is locally reversed by Zn substitution,
and that the islands are antiferromagnetic. Also, we find that the Zn impurity
locally induces a breaking of the D$_4$ symmetry. Cluster and DFT calculations
show that the D$_4$ symmetry breaking is due to the same partial lifting of
degeneracy of the nearest-neighbor oxygen sites as in the LTT transition in
La$_{2-x}$Ba$_x$CuO$_4$, similarly well-known to strongly suppress
superconductivity. These results show that in-plane oxygen $2p^5$ orbital
configurations are principally involved in the metallicity and
superconductivity of all high-T$_c$ cuprates, and provide a qualitative
symmetry-based constraint on the SC mechanism.

\end{abstract}

%\maketitle

\section{Introduction}

High-temperature superconductivity (SC) in the cuprate perovskites is still
unexplained. They are ionic crystals with a considerable variation in
composition and structure, which metallize and eventually superconduct upon
doping. Their complexity makes it difficult to find properties which are
universally significant for the SC mechanism. Perhaps the most important
and ubiquitous structural feature of the cuprates are copper-oxide (CuO$_2$)
planes, where metallization occurs. Yet the wave functions of the
superconducting charge carriers in the planes remain enigmatic, and have been
the object of intense speculation. The doping mechanism --- how impurities
introduce these carriers into the CuO$_2$ planes in the first place --- has
received comparatively much less attention. Our aim in this work is to
gain microscopic information on both the doping mechanism and the orbital
composition of the SC carriers by varying their chemical environment,
substituting a small amount of in-plane copper with zinc.

Two effects lower the SC temperature T$_c$ in cuprates so abruptly that they
may be interfering with the SC mechanism itself. One is the low-temperature
orthorombic-to-tetragonal (LTO/LTT) transition in
La$_{15/8}$Ba$_{1/8}$CuO$_{4}$ (LBCO)~\cite{Axe89}, the other is Zn
substitution of the in-plane coppers. While the former appears only in the
La cuprates, the Zn effect is observed in all of them. NMR has
found~\cite{Mahajan00} that the Curie response appearing in underdoped
YBa$_2$Cu$_3$O$_{7-\delta}$ (YBCO) upon Zn substitution is due to the
nearest-neighbor (nn) Cu sites, showing them to be electronically different
than the bulk in-plane coppers (which carry no local spin in the SC
compositions~\cite{Mila89}). ARPES has established~\cite{Terashima08} that Zn
substitution introduces no carriers into the planes, while STM has shown large
screening oscillations around the Zn site, dubbed ``desert
islands''~\cite{Pan00}.

Upon its discovery, the Zn effect was argued~\cite{Barlingay90} to provide a
major insight into the SC orbital environment. The insulating undoped cuprates
are antiferromagnetic (AF) insulators, with coppers in magnetic Cu$^{2+}$
($3d^9$) configurations, surrounded by closed-shell O$^{2-}$ ($2p^6$)
configurations. A SC-enabling orbital transition from Cu$^{2+}$/O$^{2-}$ to
Cu$^+$/O$^-$ was conjectured~\cite{Mazumdar89} to occur around 6\% hole
doping, amounting to the donation of a hole from a copper to two oxygens. This
would close the Cu $3d$ shell to $3d^{10}$, and open the O $2p$ shells to
$2p^{5.5}$, implying an oxygen-dominated SC metal. The suggested driving
mechanism for the orbital transition was ionic (Madelung energy change),
wherein the dopands affect the crystal field without their own charges being
delocalized. This scenario is henceforth referred to as \emph{ionic doping}.
Early circumstantial evidence for the orbital transition/ionic doping was the
observation of the Cu$^+$/O$^-$ configuration in the SC compositions using
X-ray absorption spectroscopy~\cite{Bianconi87}. More generally, it was
found~\cite{Qimiao93} that including the oxygens --- explicitly via the Emery
model~\cite{Emery87}, or implicitly via a second-neighbor ($t'$)
parameter~\cite{Pavarini01} in the $t$--$J$ model~\cite{Zhang88} --- is
necessary to interpret ARPES and neutron-scattering data. Significantly, this
remains true also in electron-doped cuprates~\cite{Sunko07}, indicating a
universal feature of cuprate SC.

Alternative CuO$_2$ plane doping mechanisms have been contested on either
theoretical or experimental grounds. The metallic alloying scenario, wherein
the planes are doped with delocalized dopand holes, is contradicted by ab
initio calculations showing that the Sr hole in La$_{2-x}$Sr$_x$CuO$_4$
remains localized~\cite{Perry02}. It is similarly found that the interstitial
oxygen in La$_2$CuO$_{4+\delta}$ preferentially creates an inert peroxide,
$O_2^{2-}$, with the apical oxygen~\cite{Lee06}. A more remote possibility,
the impurity-band scenario, is excluded experimentally~\cite{Bozin05}. Thus
the original ionic doping idea~\cite{Mazumdar89,Tahir-Kheli11} was left
uncontested by default, yet without microscopic experimental corroboration. Zn
substitution of in-plane Cu provides a unique opportunity to test it: because
Zn$^{2+}$ already has a closed $3d$ shell, it has no hole to donate to the
neighbouring oxygens. It should then revert them to the parent $2p^6$ and thus
locally undo the orbital transition~\cite{Barlingay90}. Such local reversal
implies a characteristic signature --- Coulomb ``domino effect'' of charge
redistribution around the Zn site --- because the charge mismatch between the
O$^{2-}$ configurations near Zn and O$^-$ in the bulk was not expected to be
stabilized across a single Cu site~\cite{Barlingay90,Mazumdar89}. In the
present work, we provide direct evidence of the conjectured orbital transition
and charge redistribution by measuring the nuclear quadrupole resonance (NQR)
signal on the Zn site itself for the first time.

The article is organized as follows. After a summary of the methods employed
in Sect.~\ref{methods}, we present the measurements in Sect.~\ref{results}. We
find that Zn creates a large insulating cluster, extending at least as far as
the nn coppers, even in nearly optimally doped YBCO, and that it breaks the
D$_4$ (square) symmetry of the planar unit cell. These observations are
subject to a detailed analysis in Sect.~\ref{analysis}, which includes
extensive theoretical modelling and comparison with the results of other work,
including Cu NMR in Zn-substituted cuprates, scanning tunneling microscopy
(STM), and neutron scattering. Finally Sect.~\ref{conclusion} collects our
conclusions.

\section{Methods\label{methods}}

\subsection{Samples and instrumentation}

Fine (micron-sized) powder samples of nearly optimally doped YBCO were
prepared for the NQR experiments by a standard solid state synthesis with
repeated annealing. High-purity precursor materials were used (Y$_2$O$_3$, CuO
and BaO, 99.99\% from Sigma-Aldrich), and the zinc was introduced by
substituting 3\% molar of CuO with ZnO ($89.6$\% $^{67}$Zn enriched, from
Eurisotop). Samples were characterized by powder X-ray, iodometry (to
determine the amount of oxygen) and Meissner shielding measurements, to
confirm their phase purity and homogeneity. Meissner shielding shows a
superconducting transition at 57~K with a width of $\sim 2$~K, which is rather
sharp for Zn-substituted YBCO \cite{Adachi01,Itoh03}. The most sensitive
indication of the microscopic homogeneity of the samples is the small width of
the $^{67}$Zn NQR lines (shown below). Relatively large amounts of powder were
used for Zn NQR due to the small concentration of zinc in the sample and low
NQR frequencies, typically $\sim 500$~mg for each sample.

Optimally doped YBCO powders were prepared with nominally 3\% substitution of
Cu by the isotopically purified $^{67}$Zn, leading to about 4\% Cu
substitution in the planes, because Zn preferentially substitutes the in-plane
Cu upon annealing~\cite{Adachi01,Itoh03,Mazumder89}. Zn and Cu NQR
experiments were performed with a standard spin-echo sequence, using a Tecmag
Apollo NMR/NQR spectrometer and a flowing gas cryostat for sample temperature
variation. In spin-lattice relaxation measurements we employed a saturation
recovery sequence (with spin echo detection). Long data acquisition times, up
to two weeks for observation of forbidden transitions, required special
precautions to stabilize the temperature.

\subsection{Cluster DFT calculations.}

As the primary purpose of the cluster calculation was to explore the
electronic neighbourhood of the Zn substituent and its effects on the lattice
semi-quantitatively, the investigated cluster was relatively small
(Fig.~\ref{dftclust}a). In addition to the atoms shown, point charges were
placed on the positions of the nearest Ba and Y ions. For all calculations we
used the hybrid functional B3LYP and the rather simple polarized Ahlrichs
basis set~\cite{Schafer92}. The RI approximation was used
throughout~\cite{Neese09}. The employed spin multiplicity of 5 corresponded to
the copper atoms being quite strongly polarized, as in previous
calculations~\cite{Bersier05}. In atomic relaxation calculations the
furthermost in-plane and all apical oxygen atoms, and all point charges were
held fixed, with other atoms free to move. Such a procedure yielded relatively
large lattice deformations, which would probably be smaller in a more
realistic cluster. However, the gross symmetry features of the deformation are
robust with respect to calculation parameters. Due to the computational cost
of high-precision calculations, we did not undertake a quantitative comparison
with the observed NQR parameters, but this should in principle be possible. We
checked that the weak orthorhombicity inherent in YBCO does not change the
nature of the lattice deformation brought upon by Zn substitution, compared to
a perfectly tetragonal initial setting. 

\subsection{Supercell DFT calculations.}

For the ground state electronic structure calculation we used plane-wave
self-consistent field DFT code (PWscf), within the Quantum Espresso (QE)
package~\cite{Giannozzi09}, and Perdew-Burke-Ernzerhof (PBE) parametrization
of the generalized gradient approximation (GGA) for the exchange-correlation
potential~\cite{Perdew96}. The ground-state electronic density was calculated
using a $5\times5\times1$ Monkhorst-Pack special $K$-point mesh, with $6$
special points in the irreducible Brillouin zone. In the PWscf code we used
projector-augmented-wave-based pseudopotentials for La, Cu and O atoms, and we
found the energy spectrum to be convergent with a $75$~Ry plane-wave cutoff.
For the in-plane (1$\times$1) unit cell parameter we used $a=7.15696\ a.u.$
which is the experimental lattice constant commonly used in the literature.
For the unit supercell in the perpendicular direction (separation between
periodically repeated La$_2$CuO$_4$ planes) we artificially expand the layer
spacing to $L=3.5a=25.11\ a.u.$, which decreases the c-axis overlap so much
that the calculation refers to a single-layer slab, appropriate for an STM
simulation. The local density of states (LDOS) $\rho(E,{\bf r})=\sum_{n,{\bf
k}}|\psi_{n,{\bf k}}({\bf r})|^2\delta(E-E_{n,{\bf k}})$, which simulates the
STM image, is  calculated from the Kohn-Sham (KS) wave functions $\psi_{n,{\bf
k}}$ and energies $E_{n,{\bf k}}$ obtained in the ground state calculation.
The spectral function distribution for the ${\bf k}$ points along the high
symmetry $\Gamma$-$\Sigma$-Y directions of the original (1$\times$1) Brillouin
zone was calculated using a recently proposed method for supercell band
structure unfolding \cite{Popescu12}.

\section{Results\label{results}}

\subsection{NQR and NMR spectra}

\begin{figure}
%\begin{tabular}{cc}
\includegraphics[width=0.9\columnwidth]{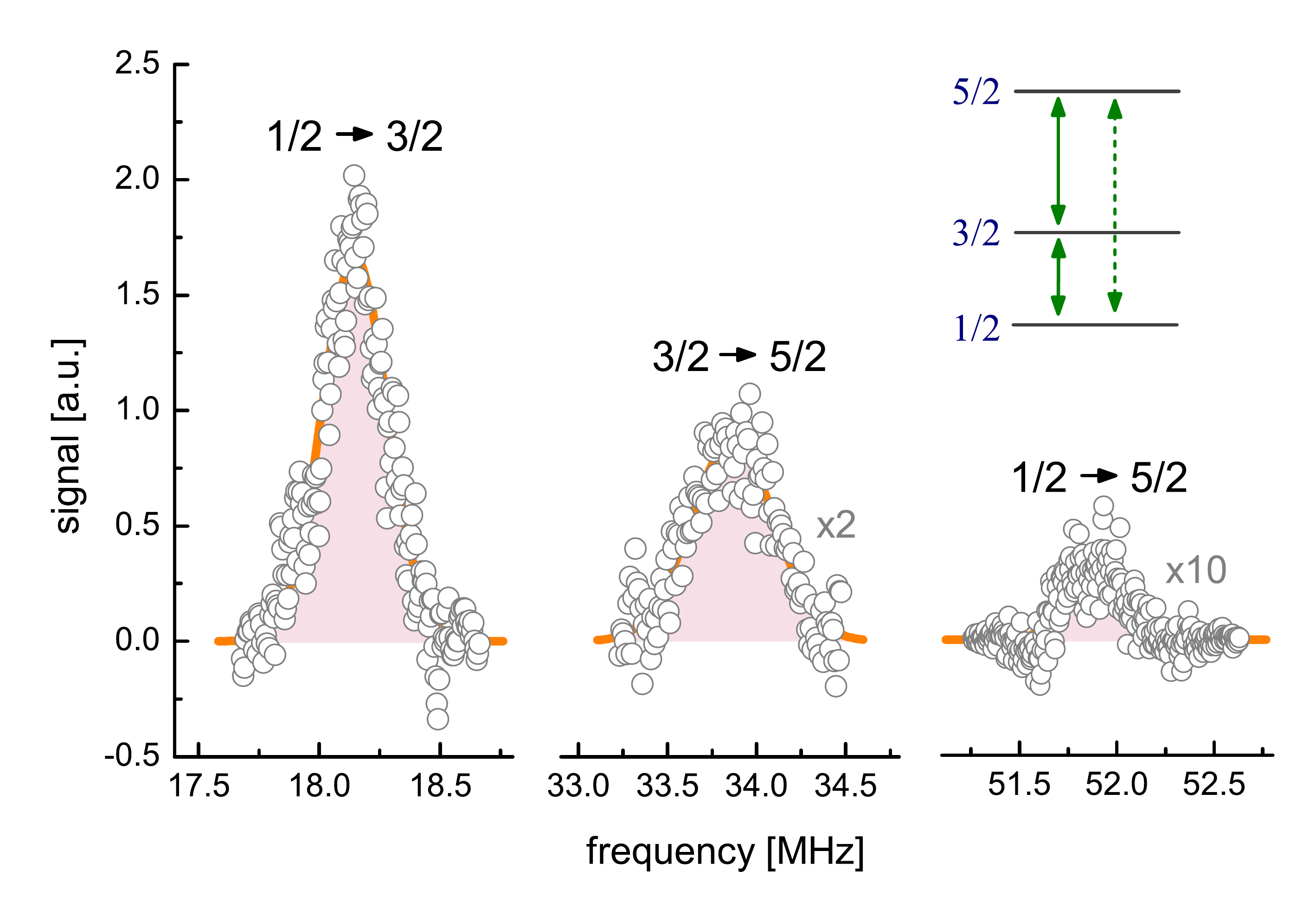}
%\includegraphics[width=0.6\columnwidth]{NMR}
%\end{tabular}
\caption{NQR signal of $^{67}$Zn (spin 5/2) in YBCO-7, at 80~K. All three
possible transitions for spin-5/2 NQR were detected, including the
``forbidden'' transition 1/2$\rightarrow$5/2, showing that the local EFG at
the zinc site is not axially symmetric. For clarity, the 3/2$\rightarrow$5/2
and 1/2$\rightarrow$5/2 intensities have been magnified by a factor of 2 and
10, respectively.}
\label{linije}
\end{figure}

Zinc nuclear magnetic resonance was chosen as a sensitive probe of the local
zinc environment in the material, and since charge effects are of interest in
this work, the principal technique is Zn nuclear quadrupole resonance (NQR).
In NQR, one detects transitions between nuclear spin levels split by local
electric field gradients, due to a coupling to the nuclear quadrupole moment.
Thus NQR provides direct information on local charge distributions. Notably,
no external magnetic field is necessary for pure NQR. We have also
successfully detected Zn nuclear magnetic resonance in external fields up to
11~T to aid in spectral line identification, but our focus here is on NQR. In
the general case of both quadrupolar and magnetic (Zeeman) splitting, the
nuclear spin Hamiltonian reads~\cite{Abragam83}
\begin{equation}
\label{hamiltonijan}
H=-\gamma \mathbf{B} \mathbf{I} + \frac{e^2QV_{zz}}{4I(2I-1)}\left[
3I_{z}^2-I^2+\eta\left( I_x^2-I_y^2\right)
\right]
\end{equation}
where $\gamma$ is the gyromagnetic ratio of the $^{67}$Zn nucleus,
$\mathbf{I}$ its spin, $Q$ its quadrupole moment, $\mathbf{B}$ the external
magnetic field, $V_{xx}$, $V_{yy}$ and $V_{zz}$ the principal components of
the electric field gradient (EFG) tensor, $\eta =
\left|\left(V_{xx}-V_{yy}\right) /V_{zz}\right|$, and $I = 5/2$ for $^{67}$Zn.
In the case of pure NQR, $\mathbf{B} = 0$ and the analysis of
(\ref{hamiltonijan}) simplifies considerably. When the EFG tensor is axially
symmetric ($\eta=0$), the Hamiltonian gives rise to two NQR lines,
corresponding to transitions 1/2$\rightarrow$3/2 and 3/2$\rightarrow$5/2 with
frequencies $\nu_{3/2 \rightarrow 5/2} = 2\nu_{1/2 \rightarrow 3/2}$ (the
negative spin levels are degenerate). In the opposite limit of $\eta \approx
1$, the two lines merge into one. YBCO possesses two distinct copper sites ---
chain Cu and in-plane Cu --- which can both in principle be occupied by the
substituting Zn. It is well known from previous NMM/NQR studies that $\eta
\approx 0$ for in-plane Cu and $\eta \approx 1$ for chain Cu sites. Thus
\textit{a priori} one would expect the Zn NQR spectrum to have two sets of
lines, one with $\eta \approx 0$ (i.e. two equally spaced lines) and one with
$\eta \approx 1$ (i.e. a single line). Yet in the experiment only one set is
observed (Fig.~\ref{linije}), and it clearly breaks axial symmetry: the two
strong lines are unequally spaced, and the``forbidden'' line corresponding to
1/2$\rightarrow$5/2 can be detected. Detection of all three lines in the
correct frequency and intensity ratios proves that the signal is due to
$^{67}$Zn, as there are no other spin-5/2 nuclei in the system. 

\begin{figure}
%\begin{tabular}{cc}
%\includegraphics[width=0.6\columnwidth]{spekt_77K}&
\includegraphics[width=0.9\columnwidth]{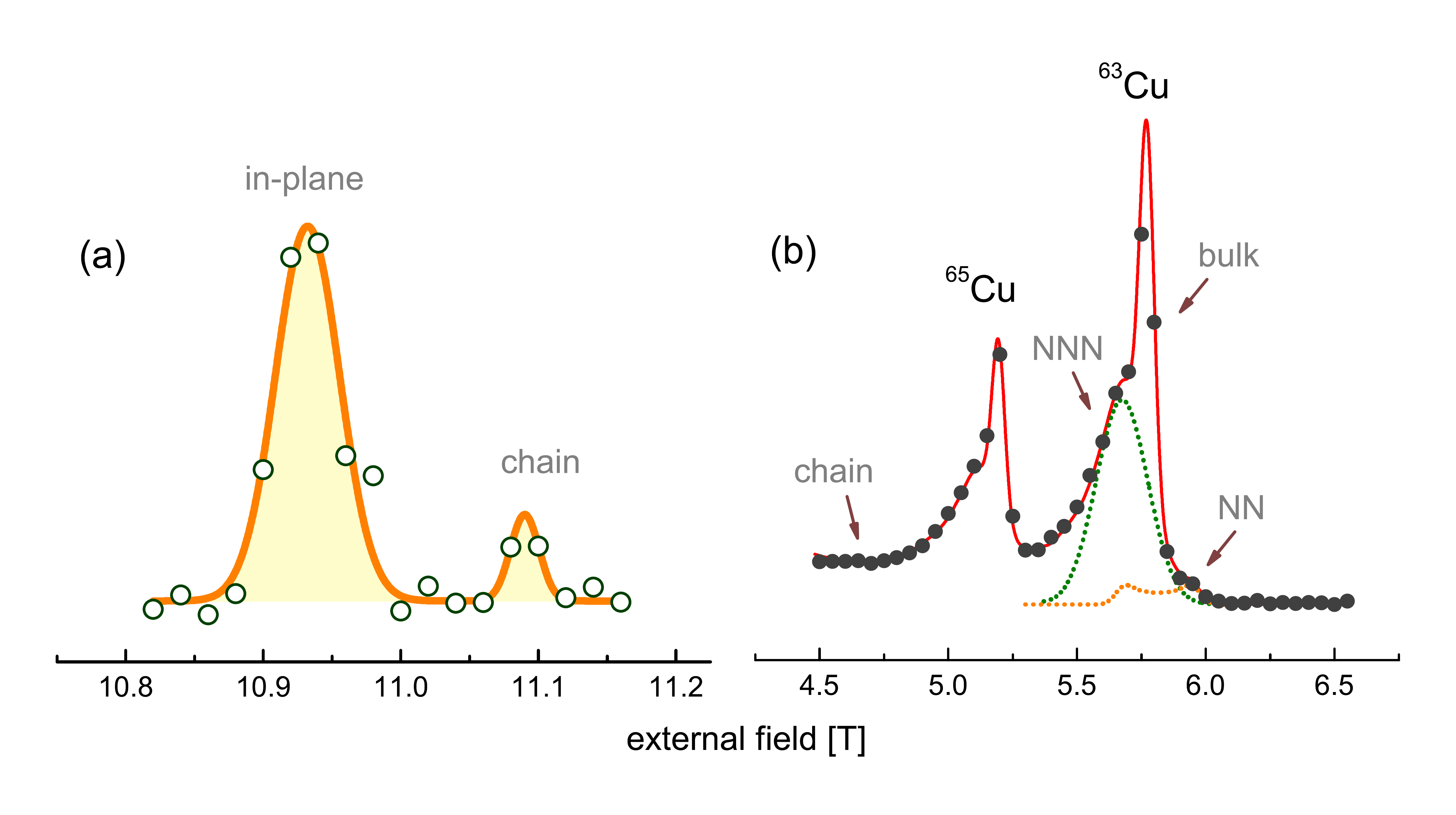}
%\end{tabular}
\caption{a) Field-sweep measurement of Zn NMR, at a frequency of 29.34~MHz and
70~K. Two lines are visible, with the weaker signal most probably originating
from in-chain Zn impurities. Inset: section of Cu NMR field-sweep spectrum
measured in an oriented powder at a frequency of 30.46~MHz and at 80~K. b)
High-field part of field sweep NMR spectrum of copper in YBCO-Zn. Two
contributions other than the bulk line are plotted as dotted lines. The fit
was performed consistently on both $^{63}$Cu and $^{65}$Cu lines, taking into
account the different gyromagnetic factors and quadrupolar moments. Due to
different symmetry of the nn line ($\eta \approx 0.2$ in the quadrupolar
Hamiltonian given in Eq.~(\ref{hamiltonijan}), compared to $\eta \approx 0.03$
for bulk and further neighbour signal) it is partially resolved in NMR, but
completely overlapped by the bulk signal in pure NQR. The offset on the
low-field side is due to chain Cu.}
\label{znnmr}
\end{figure}

However, one must still decide whether the observed signal in
Fig.~\ref{linije} originates from chain or in-plane Zn sites. Here an
additional NMR experiment with external field is helpful. In an oriented
powder, the full Zn NMR Hamiltonian (\ref{hamiltonijan}) yields a strong
central line at $\nu \approx \gamma B$ and two pairs of quadrupolar satellite
lines (which are strongly broadened if $\eta > 0$). Two relatively sharp and
distinct central lines are observed in Zn NMR at 11~T (Fig.~\ref{znnmr}a),
with the faint signal most likely originating from chain sites. Since Zn
chemically prefers to substitute for Cu in the planes, the amount of Zn in the
chains is smaller than the nominal 25\% of all Zn impurities, and thus the
chain signal is more than 5 times weaker than the in-plane Zn NMR.
Importantly, no sharp satellite lines are observed, implying that both
in-plane and chain Zn have $\eta > 0$. Broadened satellites for $\eta > 0.2$
were not reliably detectable because of great experimental difficulties in Zn
NMR. Namely, due to small Zn $\gamma$ and large $Q$, the quadrupolar and
Zeeman contributions to the NMR signal are comparable at $\sim 10$~T, making
the satellite lines spread out over a large frequency/field range in an
oriented powder. NMR in an unoriented powder was attempted in order to see the
quadrupolar effects more clearly, but the great spread of the line and
uncontrollable partial orientation of the powder in 11~T fields made the
results quantitatively unreliable. Yet qualitatively we are certain that the
in-plane signal has $\eta$ significantly greater than zero, and consistent
with $\eta = 0.25$ as seen in pure NQR.

Due to the symmetry of the chain environment, the chain sites must by contrast
have $\eta$ close to one, leading to the conclusion that the NQR lines in
Fig.~\ref{linije} are from in-plane Zn. We emphasize that the difference of
$\eta = 0.25$ and $\eta \approx 1$ is very large in terms of the EFG tensor
components $V_{\alpha \alpha}$. An extensive search for chain Zn NQR was
undertaken from $\sim 8.6$~MHz up to $\sim 45$~MHz without results. This
implies that the chain NQR signal is either too small to be observed at all,
or in the low-frequency range below 8~MHz (as would be expected from a
comparison to the equivalent Cu EFG components). The combined arguments of
NMR, site symmetry and preferential substitution thus leave no doubt as to the
detected NQR originating from in-plane substituted Zn.

The resonance lines barely move with temperature, indicating that the
asymmetry is a property of the ground state. Thus all components of the EFG
tensor at the zinc site are determined from the $^{67}$Zn NQR spectrum alone.
Using the known value of $Q$, we obtain
$\left(V_{zz},\:V_{xx},\:V_{yy}\right)=\left(-50.4,\:31.5,\:18.9
\right)\cdot10^{21}$~V/m$^2$, yielding $\eta=0.25$. These values are in
stark contrast to those at the in-plane Cu sites: the magnitude of $V_{zz}$ at
the Zn site is about five times larger than at the bulk Cu site, while the
asymmetry at the latter is only $\eta \approx 0.03$. Therefore the small
intrinsic orthorhombicity of YBCO~\cite{Pennington89}, related to the
appearance of CuO chains, cannot be the cause of the large NQR asymmetry
observed at the Zn site --- if it were, one would expect a similar D$_4$
symmetry breaking at the planar Cu sites as well.  We conclude that Zn
substitution causes a large local lattice distortion. Also, the charge
environments of the Zn and bulk plane Cu sites are obviously very different.
Previous cluster calculations~\cite{Bersier05,Kaplan02} found that the net
charge at the Zn site was indeed larger than on a bulk copper site, without
considering lattice distortions.

In addition to Zn NQR/NMR, copper sites in the neighbourhood of the Zn
impurities are of interest, providing insight into the range of electronic
effects at the Zn site itself. We have performed field-sweep Cu NMR to
identify the nearest-neighbour (nn) signal and separate it from other
further-neighbour contributions. The relevant part of the field-sweep spectrum
is shown in Fig.~\ref{znnmr}b. The two sharp peaks are bulk $^{65}$Cu and
$^{63}$Cu quadrupolar satellites; central lines are not shown for clarity. The
positions of the bulk peaks agree precisely with pure NQR of the unsubstituted
compound, and with previous measurements. Qualitatively, four distinct
contributions can be glimpsed in the spectrum, which were taken to be due to
chain Cu (offset on low-field side), bulk in-plane Cu (sharp peaks), further
Zn neighbour Cu (shoulder on low-field side of bulk Cu line) and
nearest-neighbour Cu  (shoulder on high-field side of the bulk signal). The
contributions were quantitatively separated using the following assumptions
and constraints: simultaneous treatment of both Cu isotopes, with known
intensity, gyromagnetic factor and quadrupolar moment ratios; known
quadrupolar coupling constants for bulk, further Zn neighbour and chain Cu
sites from NQR; $\eta > 0$ for the nearest-neighbour signal (since it is not
resolved in pure NQR, implying a $\nu_Q$ similar to the bulk site). Bulk and
further neighbour lines were modeled as symmetric Gaussian shapes, while nn
and chain contributions were calculated from an exact diagonalization of the
NMR Hamiltonian (1) with spin $3/2$.  and oriented powder averaging. The
asymmetry of the nn Cu line turns out to be $\eta \approx 0.2$, and its
$\nu_Q$ within 1\% of the bulk Cu value. Thus the bulk and nn Cu lines
completely overlap in pure NQR, while they are partially resolved in the NMR
spectrum. To our knowledge this is the first observation of the true nn line,
since in previous work~\cite{Itoh03,Williams01} the nnn signal (shoulder to
the left of the bulk peak in Fig.~\ref{znnmr}b, appears the same in pure NQR)
was taken to be from nn Cu on account of intensity ratios. The finite $\eta$
at the nn Cu site arises, as expected from Zn NQR, due to the lattice
distortion brought upon by Zn.

\subsection{Spin-lattice relaxation rates}

\begin{figure}
\includegraphics[width=\columnwidth]{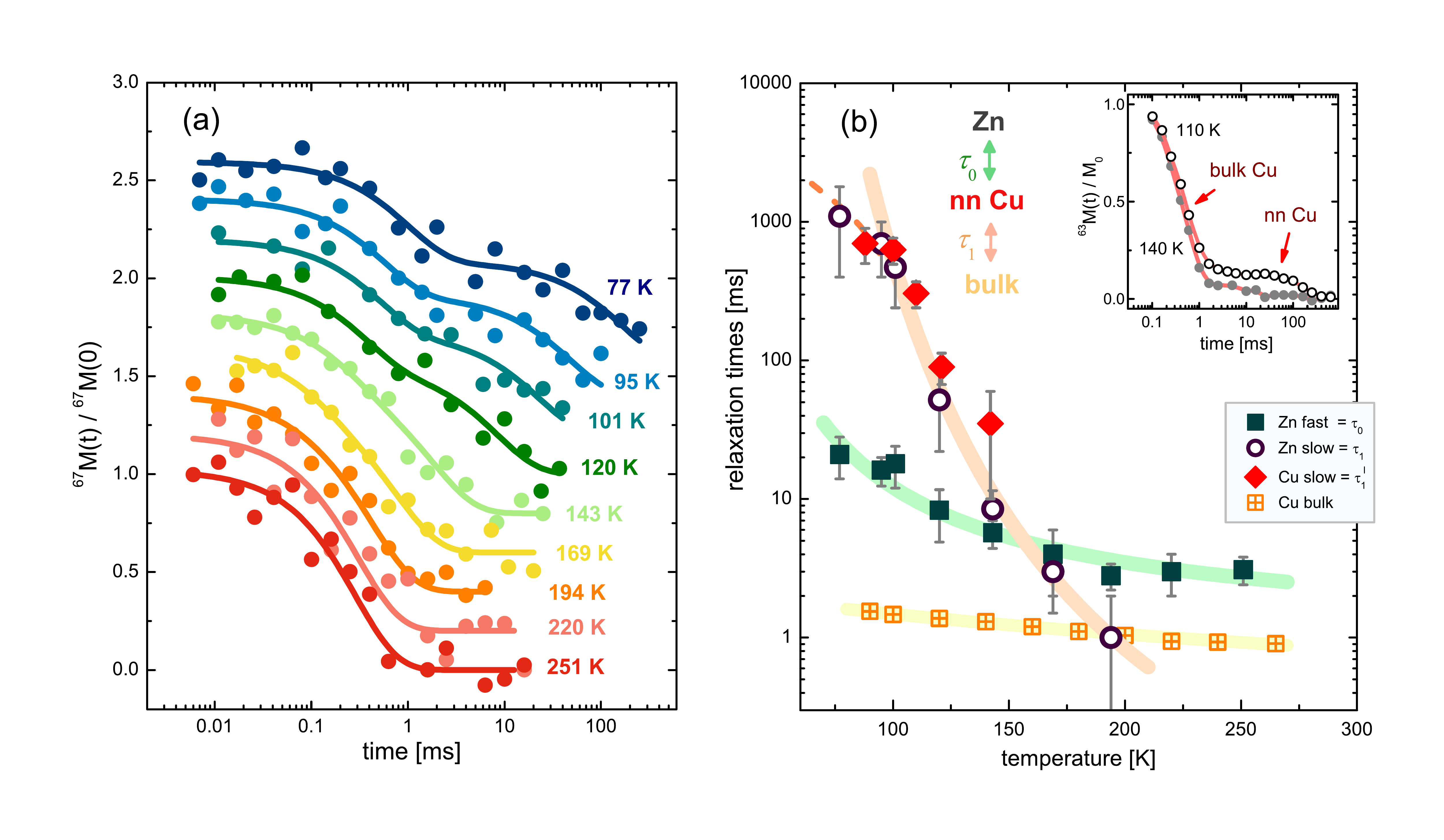}
\caption{a) $^{67}$Zn spin-lattice relaxation curves for all measured
temperatures, with two-component relaxation fits, Eq.~\ref{relax2}. The curves
are offset for clarity in steps of 0.2, with the highest-temperature
measurement unshifted. The different temperature dependences of the fast and
slow relaxation components are clearly visible. b) Spin-lattice relaxation
times of the $^{67}$Zn NQR 1/2$\rightarrow$3/2 transition, nn and bulk
$^{63}$Cu NQR, in dependence on temperature. Two components of the Zn
relaxation are plotted (full squares and empty circles), obtained from a
double relaxation model (see text). Thick full lines are Arrhenius fits for
the primary and secondary relaxation. The thin dashed line is a proposed
two-phonon process correction for the secondary relaxation (see text). The
slow copper relaxation process (full diamonds) is seen to follow the slow Zn
relaxation. Bulk Cu relaxation (crossed squares) is significantly faster than
all others and follows a ``nearly AF'' temperature dependence [line,
Eq.~(\ref{nearaf})]. Pure Curie-Weiss behaviour would be a horizontal line.
All raw relaxation times are rescaled to the standard definition of $T_1$ with
the appropriate phase-space factors~\cite{Walstedt08}: 10 for $^{67}$Zn and 3
for $^{63}$Cu. Inset: spin-lattice relaxation measurements of copper NQR at
31.5~MHz, which also contain a slow process at low temperatures. The large,
quickly decaying contribution is of bulk copper.}
\label{T1}
\end{figure}

Temperature-dependent spin-lattice relaxation measurements on $^{67}$Zn
provide further evidence of the peculiar electronic surroundings of the Zn
impurity. The first important observation is that the Zn relaxation is slow,
and involves a two-step process at temperatures below $\sim  200$~K
(Fig.~\ref{T1}a). Although the data also allow a stretched-exponential fit,
which would indicate local disorder as e.g. in CDW
systems~\cite{Dubson86,Phillips96}, we can rule that out: even a small amount
of disorder broadens the NQR line significantly (in our sample 3\% of Zn
broadens the bulk-Cu NQR line by a factor $\sim 5$.) Yet the relative width of
the Zn lines in Fig.~\ref{linije} is barely larger than the width of the Cu
NQR line in the \emph{unsubstituted} material, implying that the local
environments of the in-plane Zn impurities are uniform. This is corroborated
by the sharp onset of Meissner effect. Conversely, if
some Zn atoms did have different environments, their NQR lines would be
strongly shifted and thus undetectable in the present measurement. We conclude
that the two-step relaxation is an intrinsic property of every single Zn site,
and quantify it by a simple model. The Zn nuclear moments are insulated from
the planar metal, and feed exclusively a \emph{local} magnetization reservoir
--- which we will show is consistent with localized electronic states --- with
an ``internal'' rate $1/\tau_0$. The local states are assumed to equilibrate
quickly with nn Cu nuclear moments, which in turn relax slowly to the wider
environment with  an ``external'' rate $1/\tau_1$ (schematic on Fig.
\ref{T1}). The magnetizations $^{67}M$ and $^{63}M$ of the Zn and nn Cu
moments then evolve according to
\begin{equation}
\label{relax1}
\begin{array}{rcl} \frac{d {^{67}M}}{dt} &=& -
\frac{1}{\tau_0}({^{67}M}-{^{63}M}) \\ \frac{d {^{63}M}}{dt} &=&
-\frac{1}{\tau_1}{^{63}M}-\frac{1}{\tau_0}({^{63}M}-{^{67}M}). \end{array}
\end{equation}
Solving the coupled system gives the $^{67}$Zn magnetization,
\begin{equation} \frac{^{67}M(t)}{^{67}M_0} = A_+ e^{-\alpha_+ t} + A_-
e^{-\alpha_- t}. \label{relax2}
\end{equation}
with $\alpha_\pm = 1/\tau_0+1/2\tau_1 \pm \sqrt{1/\tau_0^2+1/4\tau_1^2}$ and
$A_\pm = \frac{1}{2}\left(1\mp 1/\sqrt{1+4\tau_1^2/\tau_0^2} \right)$. Such a
coupled double-exponential relaxation fits the experimental curves well (inset
to Fig.~\ref{T1}), up to $T \approx 200$~K where $\tau_1$ becomes too small to
be resolved. The constants $A_\pm$ are wholly given in terms of the two
relaxation times $\tau_0$ and $\tau_1$, without introducing additional fitting
parameters. The fit thus has the same number of parameters as a stretched
exponential, and yields smaller overall least-square deviations than
stretched-exponential fits at low temperatures. A somewhat similar relaxation
model was proposed for nn Cu in Zn-Y248~\cite{Williams01}.

Additional corroboration of the two-step relaxation model can be obtained from
nn Cu spin-lattice relaxation measurements. The nn Cu signal was identified
with the help of field-sweep NMR, as explained in detail above. Unfortunately,
the nn Cu and bulk Cu lines overlap in pure NQR; this is also perhaps the
reason why the true nn signal was overlooked in previous work, with the
next-nearest neighbors possibly being confused with the nearest
ones~\cite{Itoh03,Williams01}. Nevertheless, a long tail due to nn Cu was
detected in the NQR relaxation curves (Fig.~\ref{T1}b inset), with the values
of the relaxation times $\tau_1'$ agreeing with $\tau_1$ from Zn measurements.

The finding of a slow nn Cu relaxation in agreement with the slow Zn process
and the large difference between $\tau_0$ and $\tau_1$ at low temepratures
constrain our interpretation by excluding several simple alternatives. The
second, 'external' process has to be the slower one, or the two stages would
not be resolved. The first magnetization reservoir has to feed the second
exclusively, otherwise the specific algebraic relationships $2\alpha_-\approx
1/\tau_1$ and $\alpha_+\approx 2/\tau_0$, valid when $\tau_0\ll\tau_1$, would
change and spoil the rescaling of raw data, required for the extracted Zn
$\tau_1(T)$ to coincide with the directly measured Cu $\tau_1'(T)$. Finally,
the intermediate states, by which Zn moments relax to the nn Cu moments, do
not thermalize, or otherwise a third magnetization reservoir would be
necessary to describe the data. In the limit of long $\tau_1\gg\tau_0$, $A_\pm
\rightarrow 1/2$, in agreement with the experimental relaxation curves shown
in Fig.~\ref{T1}a. This agreement eliminates a three-step process. 

The relaxation times $\tau_0$, $\tau_1$ and $\tau_1'$ are plotted in
Fig.~\ref{T1}b, with clearly very different magnitude and temperature
behaviour. Bulk Cu relaxation (which agrees with measurements in unsubstituted
YBCO~\cite{Uldry05}) is shown for comparison, and it is  qualitatively
different than Zn in the entire temperature range, in addition to being
significantly faster. This shows that the electronic fluctuations to which the
Zn impurity is exposed are not metallic, as for the bulk copper sites. A
quantitative analysis and comparison to electronic structure calculations will
make this statement more precise, and elucidate the nature of the
fluctuations.

\section{Analysis and discussion\label{analysis}}

\subsection{NQR relaxation}

Apart from the two-step relaxation process, the Zn $\tau_0$ and $\tau_1$ data
are irreconcilable with a metallic environment, either the usual (Korringa)
$1/(T_1T)\sim\mbox{\textit{const.}}$, or the ``nearly AF'' temperature
dependence~\cite{Moriya90,Monien91,Nakai10}
\begin{equation}
1/(T_1T)\sim a+b/(T+T_{AF}),
\label{nearaf}
\end{equation}
obeyed by the bulk copper $T_1$ in Fig.~\ref{T1}b. Also, previous observations
of enhanced relaxation rates and NMR shift at further neighbor copper
sites~\cite{Itoh03} and lithium impurities~\cite{Bobroff99,Ouazi04},
interpreted in terms of strong nearly-AF metallic fluctuations, are in stark
contrast to the observed low-temperature slowing down of Zn relaxation.
Therefore the immediate Zn environment is insulating, rather than a nearly-AF
metal. In line with that reasoning, the temperature dependence of the slow
component $\tau_1$ is well described by an Arrhenius curve of the form
$1/\tau_1 \sim e^{-\Delta/kT}$ with a gap of $\Delta \sim 0.1$~eV, consistent
with a superexchange interaction between the nn Cu electronic local moments.
The activated behavior implies that the cluster is insulating, while the scale
of the interaction confirms the conjecture~\cite{Mahajan00} that the cluster
around the Zn site is AF, supporting the original explanation for the observed
AF Curie law~\cite{Mahajan00}. The insulating cluster is evidently small
enough for AF fluctuations to appear as a finite-size effect, as seen in
molecular magnets~\cite{Wang11}. Such collective fluctuations are expected to
enhance the Curie response over that of individual spins.

As a fine detail, the $\tau_1$ data points in Fig.~\ref{T1} seem to deviate
from the Arrhenius curve at lowest temperatures and longest relaxation times.
Although we have only one $\tau_1$ and one $\tau_1'$ point there, the tendency
of both to the same deviation brings us to tentatively propose a correction at
the lowest temperatures, shown by a dashed line in the figure. It was modelled
by the dominant two-phonon Raman process~\cite{Abragam83}, with the
characteristic temperature dependence $1/T_{1} \sim T^2$ (which might actually
be faster due to the relevant temperatures being close to the Debye
temperature). It is not unrealistic that the transition matrix elements are
small enough to lead to relaxation times of the order of
0.1~second~\cite{Abragam83}. However, the small number of data points, at the
edge of the measurable range, precludes a positive identification.

The fast Zn relaxation component, $\tau_0$, corresponding to the 'internal'
relaxation process, is similarly satisfactorily described by an activated
temperature dependence, albeit with a much smaller gap than $\tau_1$. A fit
(solid line through $\tau_0$ points in Fig.~\ref{T1}b) yields a gap of $\sim
20$ meV, suggesting that slightly gapped local electronic states are
responsible for the relaxation. Cluster and periodic DFT calculations indicate
that just such states arise induced by Zn, and are responsible for the local
D$_4$ symmetry breaking, as follows.

\subsection{Microscopic analysis}

\begin{figure}
\begin{tabular}{ccc}
\includegraphics[width=0.35\columnwidth]{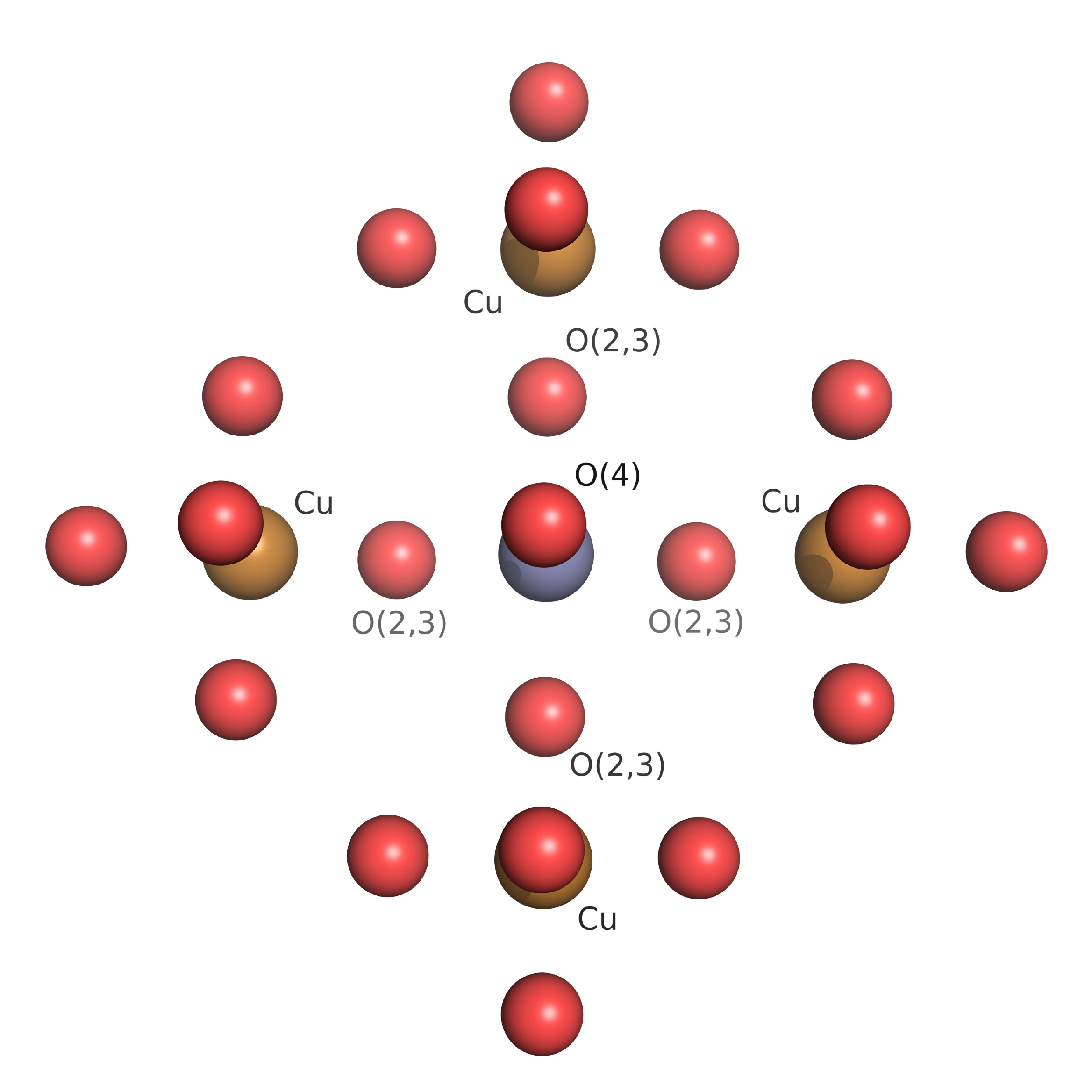}
&\includegraphics[width=0.25\columnwidth]{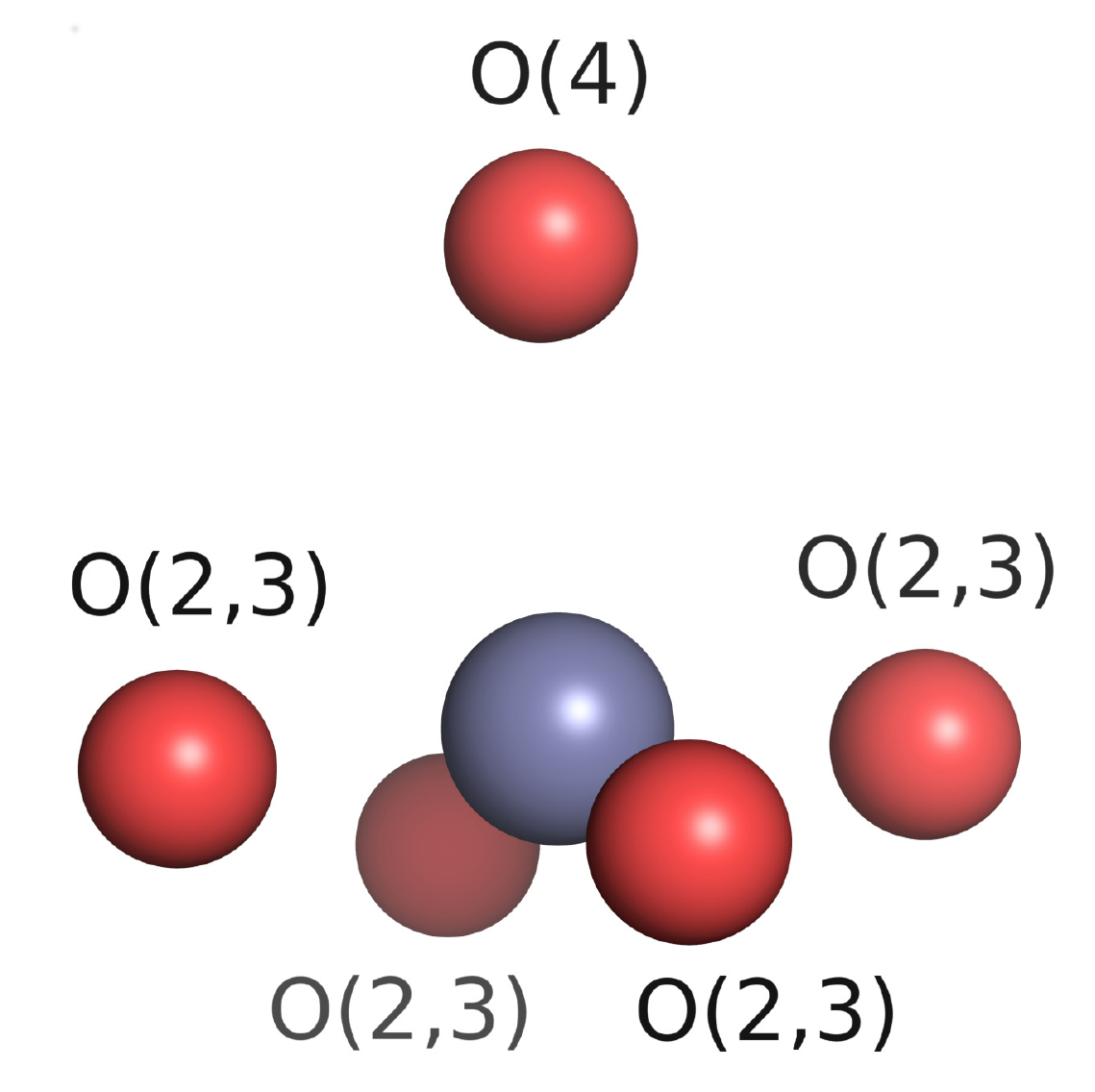}
&\includegraphics[width=0.3\columnwidth]{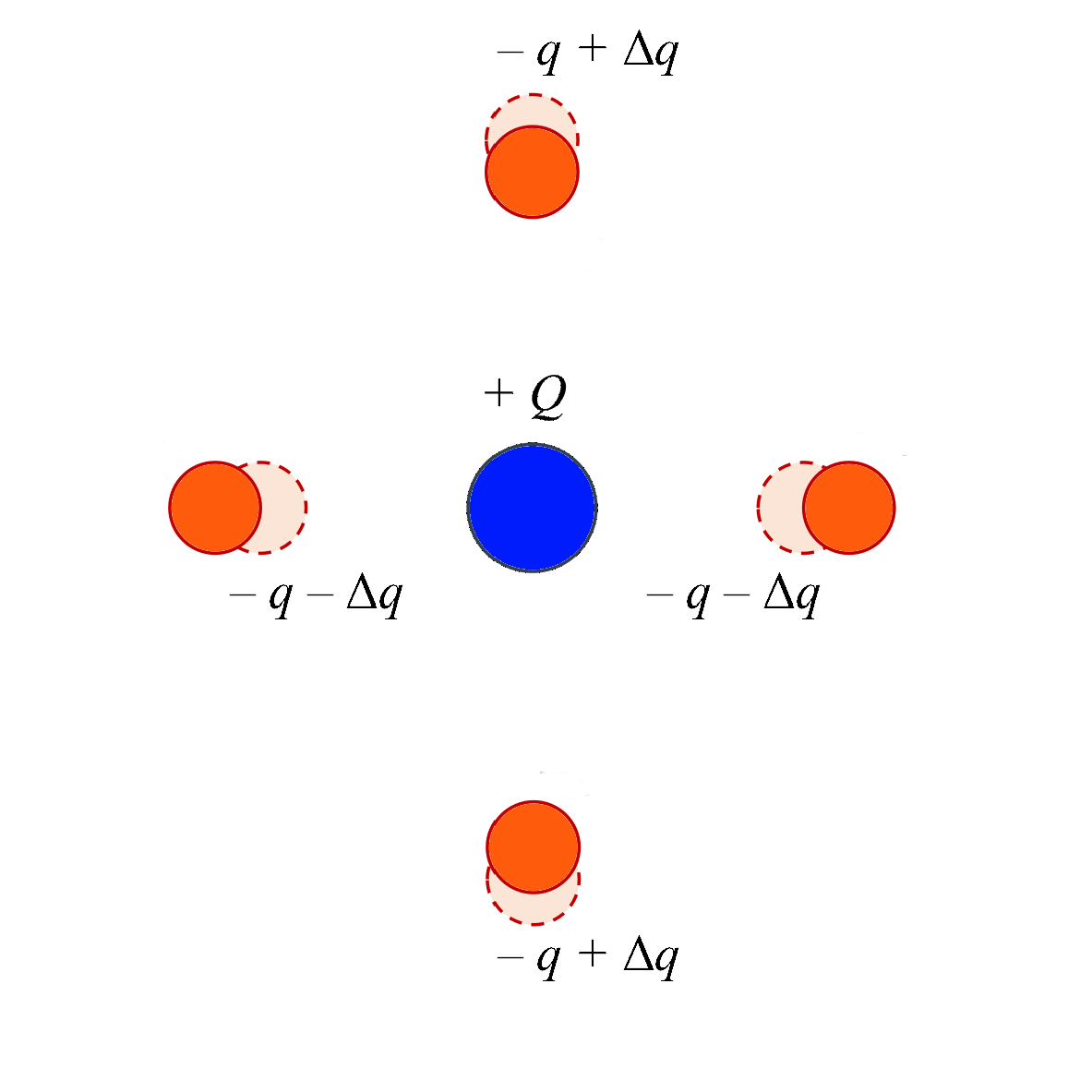}\\ (a)&(b)&(c)
\end{tabular}
\caption{(a) The entire DFT cluster (without point charges) used for
electronic structure calculations in ORCA. O(4) are apical, and O(2,3)
in-plane oxygen atoms, and the grey central atom is zinc. Realistic YBCO
lattice parameters are taken from X-ray diffraction. (b) Deformation of the
ZnO$_5$ pyramid. The two O$_{2,3}$ oxygens in the planar unit cell are no
longer equivalent. (c) Illustration of the simple 2D electrostatic argument
used in the text. The movement of ions is denoted by the dotted outlines. Here
$Q=4q$.}
\label{dftclust}
\end{figure}
The Zn 3d orbital is closed (3d$^{10}$), making it a good candidate for local
DFT calculations without strong correlations~\cite{Bersier05,Kaplan02}. We
have performed a cluster DFT calculation of molecular states in plaquettes of
four Cu atoms, a central Zn impurity, and their oxygen
pyramids~\cite{Schafer92,Neese09,Neese12}, shown in Fig~\ref{dftclust}a. We
obtain a lattice distortion with broken tetragonal symmetry, with low-lying
excited states appearing with the distortion. The D$_4$ symmetry is lost by
breaking the degeneracy between the two oxygens in the planar unit cell
(Fig.~\ref{dftclust}b). The same symmetry breaking appears in the LTT
tilt~\cite{Axe89,Barisic90}. The excited states are $\approx 100$~meV above
the ground state, which is an overestimate, given the crudeness of the
calculation.

\begin{figure}
\setlength{\tabcolsep}{10pt}
\begin{tabular}{cc}
\includegraphics[width=0.4\columnwidth]{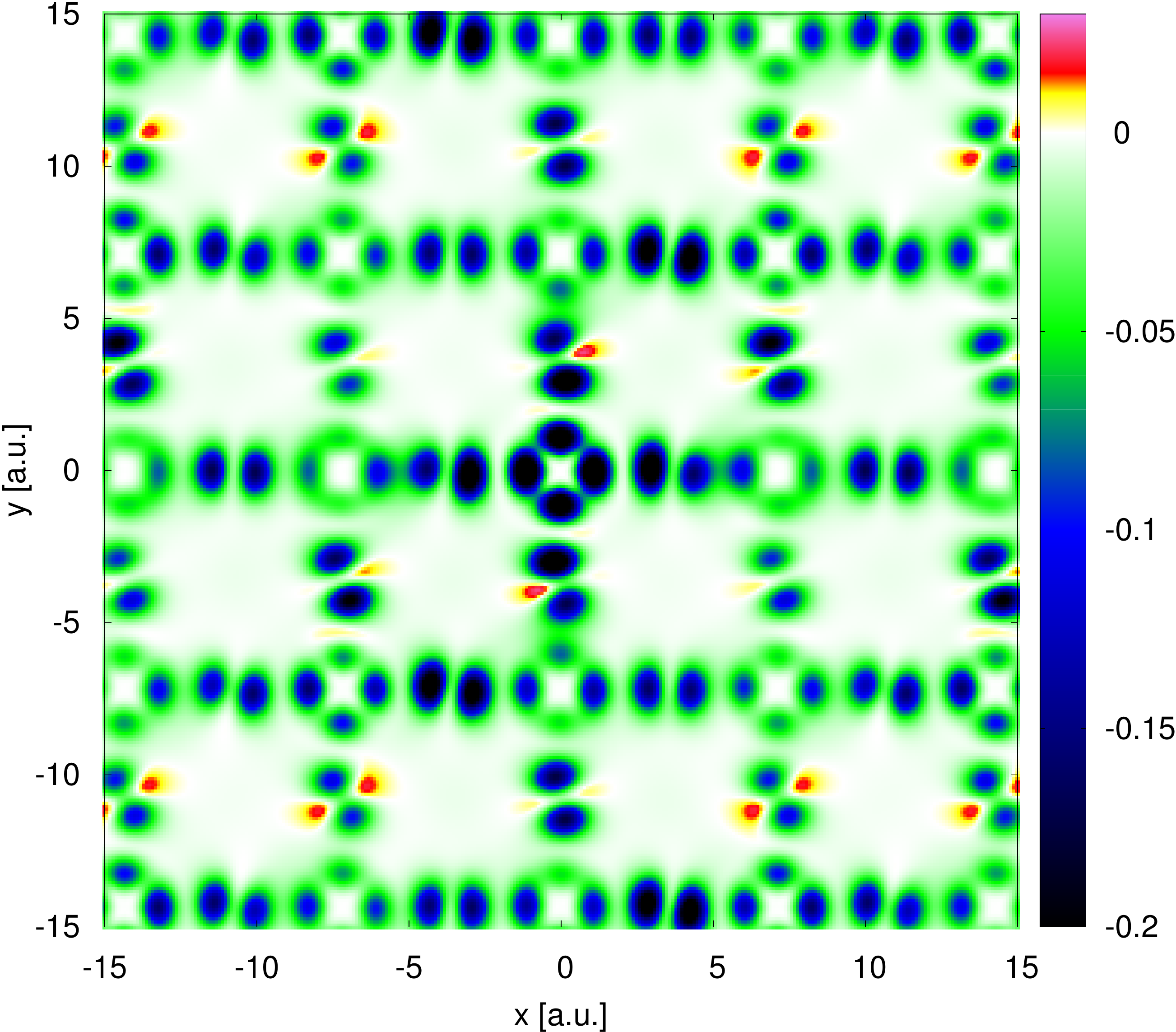}
&\includegraphics[width=0.5\columnwidth]{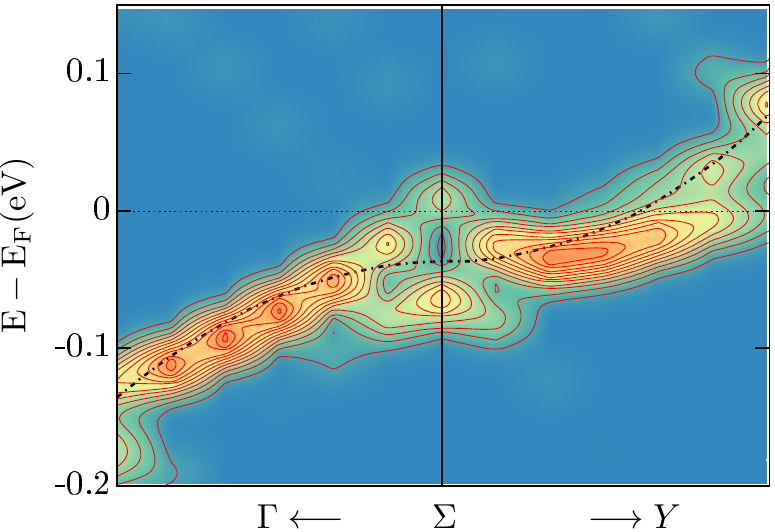}\\ (a)&(b)
\end{tabular}
\setlength{\tabcolsep}{\deftabcolsep}
\caption{(a) Difference of simulated STM images, integrated in a $20$~meV
window at $E_f$, for a slab with $3\times 3$ supercell of La$_2$CuO$_4$ unit
cells with and without the central Cu replaced by Zn. Note the different
densities in the $x$- and $y$-directions. Color scale units:
$10^{-3}a_0^{-3}$. (b) Spectral function (arbitrary units) near the van Hove
singularity for a $5\times 5$ supercell, corresponding to 4\% Zn doping in the
plane.}
\label{dftslab}
\end{figure}
We also performed periodic DFT
calculations~\cite{Giannozzi09,Perdew96,Popescu12} with supercells consisting
of $3\times 3$ and $5\times 5$ La$_2$CuO$_4$ unit cells, with central Cu
replaced by Zn. We intentionally employ the tendency of DFT calculations to
spurious metallicity, so as to model the disruption by Zn under circumstances
least favorable to it. Remarkably, Zn breaks the \emph{electronic} D$_4$
symmetry in this ordered lattice even if the atoms are held fixed in their
ideal-lattice positions, as shown in Fig.~\ref{dftslab}a. (In STM
measurements~\cite{Hamidian12} the symmetry is broken even further, which may
be a surface effect.) Breaking the degeneracy of the two oxygens in the unit
cell splits the van Hove singularity~\cite{Barisic90}. For realistic 4\% Zn
doping the excitation scale from $E_f$ drops to about $20$~meV
(Fig.~\ref{dftslab}b), which is an underestimate, because of the artificial
periodicity and lack of atomic relaxation in the model system.

Qualitatively, the electronic D$_4$ symmetry breaking in the periodic
calculation is consistent with the deformation in the small-cluster
calculation. The energy scale of the local cluster is bracketed between $20$
and $100$~meV by these two limits. Without discussing the coupling mechanism,
we conclude that the Zn moments relax to nn Cu moments via electronic states
of the planar ZnO$_4$ cluster.

The D$_4$ symmetry breaking may be understood intuitively as a simple
electrostatic effect. In a first approximation, the ions in the basic ZnO$_4$
square can be regarded as point charges (Fig.~\ref{dftclust}c). The square is
then unstable in two ways. First, a transfer of a small charge $\Delta q$
between the oxygens and simultaneous shift of the ions \textit{in the plane}
(pictured in Fig.~\ref{dftclust}c) lowers the energy and breaks the square
symmetry. The tendency towards such a transfer is corroborated by the
calculation in Fig.~\ref{dftslab}a, which shows an electronic instability
($\Delta q\neq 0$) in the plane when the atoms are \emph{not} allowed to
relax, and oxygen orbitals must accommodate the extra charge $\pm \Delta q$.
Second, even if $\Delta q = 0$, the square is still unstable towards a shift
of the oxygens \textit{out of the plane}, with one oxygen in the unit cell
going up, and the other down.  This instability becomes relevant when all Zn
and O orbitals are closed --- as they are in the present case --- making
charge transfer impossible. Such a shift in itself keeps the mirror symmetry
of the plane intact, but once the apical oxygen (and other ions) are included,
the mirror symmetry is broken and the lowest energy state is reached by
oxygens shifting unequally both out of the plane and towards/away from the Zn.
This is seen in the full relaxed 3D cluster calculation, Fig.~\ref{dftclust}b.
Importantly, because the Zn impurity is highly charged in comparison to its
surroundings, long-range contributions from far-away ions cannot remove the
Coulomb instability as they do in the bulk; similarly, entropy contributions
to the free energy are not important for the local effects of Zn. Also, closed
Zn and O orbitals mean that the point-charge approximation is expected to be
quite good. We conclude that the D$_4$ symmetry breaking occurs because the Zn
impurity triggers elementary electrostatic instabilities in a local
point-charge (closed-orbital) configuration. 

Our observation that the Zn cluster is insulating is in accord with the
observation of a zero-bias peak at the Zn site in STM~\cite{Pan00}. We concur
with Ref.~\cite{Pan00} that the peak is a scattering resonance, i.e.
interference peak in the Friedel screening. An interference maximum is not
surprising at an insulating site, like at the position of a hole in an antenna
array. For Zn-BSCCO, this picture has been corroborated by a concrete
calculation~\cite{Wang05}. The question is how big the insulating hole is, and
we find that it includes the nn coppers, where STM~\cite{Pan00} finds zero
DOS. As also noted in Ref.~\cite{Pan00}, the appearance of metallic screening
at zero bias is direct evidence that SC is destroyed around the Zn site,
because the STM measurements were carried out deep below T$_c$.

\subsection{Inferences for collective states}

The present results have important repercussions for the composition of the SC
metal in the cuprates. The gapped NQR relaxation of the Zn and nn Cu sites,
and the large EFG's at the Zn site, are incompatible with a metallic
environment. However, an insulating environment indicates that the in-plane
oxygen atoms in the SC materials are in a different orbital configuration than
in the parent materials~\cite{Barlingay90}. One can express the ionic doping
proposal~\cite{Mazumdar89} as a chemical reaction in the solid state,
\begin{equation}
\mbox{Cu$^{2+}+$O$^{2-}\rightleftharpoons\;$Cu$^{+}+$O$^{-}$},
\label{eqorbtrans}
\end{equation}
whose balance is tuned by doping. As mentioned above, the right-hand-side
ionic configuration was observed very early~\cite{Bianconi87} to appear
concomitantly with SC. We observe the insulating islands precisely where the O
$2p^5$ (O$^{-}$) configuration of the SC metal \emph{cannot} appear, because
Zn is already in the $3d^{10}$ (Zn$^{2+}$) configuration next to the parent
$2p^6$ (O$^{2-}$). The islands are in the original (parent) AF Mott insulator
configuration. Because the parent $3d^9$ (Cu$^{2+}$) configuration is
converted to $3d^{10}$ (Cu$^+$) in the SC compositions, the on-site repulsion
$U_d$, due to the large ionization energy of the $3d^8$ (Cu$^{3+}$)
configuration, is a second-order effect in the normal state of the SC metal at
low energy. Indeed, a careful dynamical treatment of $U_d$ in the limit of
significant Cu--O covalency, but starting from the $3d^{10}$ vacuum, shows
that the direct effects of $U_d$ are observed in ARPES away from the Fermi
energy~\cite{OSBarisic12}.

As described so far, the observed local reversal of the
reaction~(\ref{eqorbtrans}) to the left-hand-side would reduce the SC
fraction, but not necessarily the SC T$_c$. However, we have uncovered an
additional unsuspected effect of the Zn impurities on the SC metal itself. The
Coulomb domino effect around Zn sites breaks D$_4$ symmetry and lifts the
degeneracy of the oxygens in the planes, in the same manner as the LTT tilt in
LBCO~\cite{Axe89,Barisic90}. In Fig.~\ref{dftslab}a, it progresses to the
next-nn O sites beyond the nn Cu sites (consistently with the fact that the nn
Cu sites are not metallic either), involving by extension the next-nn Cu sites
on their other side. This observation elevates the O$_x$--O$_y$ site-energy
splitting to a universal antagonist of SC in the cuprates. Its microscopic
effect on the SC is obscured in the LTT tilt, because the split van Hove
singularity shifts electrons away from the Fermi surface~\cite{Barisic90}. Any
lowering of the density of states at the Fermi surface should lower T$_c$,
whatever the SC mechanism. We find, however, that scattering on the
oxygen-splitting phonon is negatively affecting the SC electrons
\emph{microscopically} in all cuprates, given that all are subject to the Zn
effect.

The requirement that the two oxygens in the unit cell be degenerate is a
qualitative constraint on the SC mechanism. It means that the oxygen orbitals
in the superconducting metal must appear in symmetry-determined, not
parameter-determined, superpositions, because only the former are singularly
perturbed by even a small O$_x$--O$_y$ splitting. One such oxygen singlet
superposition was found~\cite{Sunko09} to destabilize the Zhang-Rice
singlet~\cite{Zhang88}, provided the Cu--O overlap is effectively reduced in
favor of the O$_x$--O$_y$ overlap. Such a reduction appears both in the
standard slave-particle scenarios starting from the $3d^{9}$
vacuum~\cite{Qimiao93}, and if $3d^{10}$ is the relevant
vacuum~\cite{OSBarisic12}, as inferred here.

As an additionial point, our measurements suggest an explanation of the
effects of Zn doping on the LTT phase in LBCO~\cite{Hucker11a}. The
charge-stripe correlations disappear because they are disrupted by the
insulating islands. The spin-stripe response is enhanced because the LTT tilt
is stabilized by the similarly-acting Zn. Like in the local Curie response
induced by the same Zn, magnetic responses are again enhanced by the
non-magnetic Zn impurity because it acts strongly in the charge channel ---
creating an insulator, and deforming the lattice, respectively.

\section{Conclusion\label{conclusion}}

We have measured the NQR signal in Zn-substituted optimally doped YBCO on the
Zn sites themselves for the first time. We find that Zn creates large
insulating islands. Instead of the metal shielding the Zn impurity, the
impurity reverses the metallic vacuum, so that the metal disappears. Locally,
the optimally doped material reverts to the closed O $2p^6$ orbitals
characteristic of the parent compound, in which the Cu $3d^9$ orbitals are
open, and magnetic. Conversely, the SC metal vacuum is based on open O $2p^5$
orbitals and closed Cu $3d^{10}$ orbitals. Zn impurities are experimentally
observed to break the D$_4$ symmetry of the lattice site which they occupy.
Calculations show this occurs by breaking the degeneracy of the two planar
oxygen orbitals within a unit cell, which is thus found to be strongly
detrimental to SC in all cuprates, pointing to the direct involvement of
O$_x$--O$_y$ degenerate superpositions in the SC mechanism.

\section{Acknowledgments}

Conversations with S.~Bari\v si\'c, A.~Dul\v ci\'c, P. Lazi\'c, S.~Mazumdar,
and D.~Pavuna are gratefully acknowledged, as well as the help of
T.~Cvitani\'c with sample preparation. D.K.S. thanks the organizers for an
invitation and a stimulating time at the ECRYS-2011 Workshop on Electronic
Crystals.

This work was supported by the Croatian Government under project
No.~119-1191458-0512, by the University of Zagreb grant No.~202301-202353,
and by the Croatian Science Foundation (HRZZ) Grant 2729.

\section*{References}

%\begin{thebibliography}{11}

\providecommand{\newblock}{}

%\end{thebibliography}
\end{document}